\documentclass[superscriptaddress,groupedaddress,nofootnoteinbib,11pt]{article}
\pdfoutput=1
\usepackage{color}
\usepackage{graphicx}
\usepackage{dcolumn}
\usepackage{bm}
\usepackage{amssymb}
\usepackage{amsmath}
\usepackage{sectsty}
\usepackage{colortbl}

\usepackage{latexsym}
\usepackage{float}
\usepackage{ifthen}
\usepackage{enumerate}
\usepackage{url}
\usepackage[force]{feynmp-auto}
\usepackage{jcappub}
    \usepackage{picinpar}
    \usepackage{colortbl}
\usepackage{multirow}
	\usepackage{float}
	      \usepackage{setspace}
\usepackage{array}
\usepackage{bm}
\usepackage{amsopn}
\usepackage{tablefootnote}
\renewcommand{\vec}[1]{\bm{\mathrm{{#1}}}}
\usepackage{booktabs}
\usepackage[table]{xcolor}
\definecolor{lightgray}{gray}{0.9}

\DeclareMathOperator{\Or}{O}

\def\d{\textrm{d}}

\def\ba{\begin{eqnarray}}
\def\ea{\end{eqnarray}}
\def\beq{\begin{eqnarray}}
\def\eeq{\end{eqnarray}}
\def\noi{\noindent}

\def\mpl{M_{g}}

\def\L{\mathcal{L}}

\def\({\left(}
\def\){\right)}

\def\nn{\nonumber}
\def\mn{_{\mu \nu}}
\def\stu{St\"uckelberg }

\newcommand{\rhomg}{{\hat\rho_{m,g}}}
\newcommand{\rhomf}{{\hat\rho_{m,f}}}
\newcommand{\meff}{{ m_{{\rm eff}} }}

\def\mupn{^\mu_{\ \nu}}
\def\<{\langle}
\def\>{\rangle}

\def\K{\mathcal{K}}

\definecolor{verde}{rgb}{0,0.5,0}

\usepackage{titlesec}
\def\d{\textrm{d}}

\newcommand{\para}[1]{\par\vspace{2mm}\noindent\textbf{{#1}}.---}

\newcolumntype{Q}{>{$\displaystyle}l<{$}}
\newcolumntype{q}{>{\columncolor[gray]{0.9}$\displaystyle}l<{$}}
\newcolumntype{R}{>{$\displaystyle}r<{$}}
\newcolumntype{S}{>{$\displaystyle}c<{$}}
\newcolumntype{s}{>{\columncolor[gray]{0.9}$\displaystyle}c<{$}}
\newcolumntype{T}{>{\columncolor[gray]{0.9}}c<{}}

\newsavebox{\tableA}
\newsavebox{\tableB}

\newsavebox{\boxplot}
\newsavebox{\boxplota}

\newboolean{editorial}
\setboolean{editorial}{true}
\newcommand{\editorial}[2]{\ifthenelse{\boolean{editorial}}{\textcolor{red}{ [\textsf{\textbf{{#1}}}:} \textcolor{blue}{\textsf{{#2}}}\textcolor{blue}{]}}{}}

\definecolor{dullpurple}{rgb}{0.431,0.188,0.534}
\definecolor{darkgreen}{rgb}{0.133,0.545,0.133}
\definecolor{verde}{rgb}{0,0.5,0}

\newboolean{comment}
\setboolean{comment}{true}
\newcommand{\comment}[2]{\ifthenelse{\boolean{comment}}{\textcolor{blue}{{{{#1}}}: }\textcolor{verde}{{#2}}}{}\\}

\newboolean{red2}
\setboolean{red2}{true}
\newcommand{\redtwo}[2]{\ifthenelse{\boolean{red2}}{\textcolor{red}{#1}}\\}

\begin{document}

\title{
Mild bounds on bigravity from primordial gravitational waves}

\author{Matteo Fasiello$^{a,b}$ and  Raquel H. Ribeiro$^{c,a,d}$}
\affiliation{$^{a}$CERCA/Department of Physics, Case Western Reserve University, \\
10900 Euclid Ave, Cleveland, OH 44106, U.S.A.}
\affiliation{$^{b}$Stanford Institute for Theoretical Physics, Stanford University,\\ Stanford, CA 94306, U.S.A.}
\affiliation{$^{c}$School of Physics and Astronomy, Queen Mary University of London, \\
Mile End Road, London, E1 4NS, U.K.}
\affiliation{$^{d}$Perimeter Institute for Theoretical Physics, \\
31 Caroline St N, Waterloo, Ontario, N2L 6B9, Canada}

	\emailAdd{Matteorf@stanford.edu}
	\emailAdd{R.Ribeiro@qmul.ac.uk}

\abstract{
If the amplitude of primordial gravitational waves is measured in the near-future, what could it tell us about bigravity? 
To address this question, we study massive bigravity theories by focusing on a region in parameter space which is safe from known instabilities. 
Similarly to investigations on late time constraints, 
we implicitly assume there is a successful 
implementation of the Vainshtein mechanism which 
guarantees that standard cosmological evolution is largely unaffected.
We find that 
viable bigravity models 
are subject to far less stringent constraints 
than massive gravity, where there is only one set of (massive) tensor modes. 
In principle sensitive to the effective graviton mass at the time of recombination, 
we find that in our setup the
primordial tensor spectrum is more responsive to the dynamics of the massless tensor sector rather than its massive counterpart. 
We further show there are intriguing windows in the parameter space of the theory which could potentially
induce distinct signatures in the $B$-modes spectrum.
}

\maketitle

	\section{Introduction}
	\label{sec:introduction}

	It is well known that 
	the pioneering work of Fierz and Pauli~\cite{Fierz:1939ix} sparked a theoretical programme 
	aimed at deriving a consistent 
	fully non-linear theory of a massive spin-2 field.
	This search has been refueled by the 
	observation of the current accelerated 
	expansion of the universe~ \cite{Perlmutter:1997zf,Riess:1998cb,Tonry:2003zg}, an acceleration which is sometimes predicted in massive gravity models.
	There were, however, serious 
	obstructions to constructing such theories, both theoretical and observational. Of these,
	the presence of the Boulware--Deser ghost~\cite{Boulware:1973my} and the vDVZ discontinuity 
	which prevented the massless limit 
	from reproducing the results 
	of General Relativity (GR) in the regime where they have been verified, are the most notorious.

	The ghost-free extension was put forward 
	only very recently~\cite{deRham:2010gu,deRham:2010ik,deRham:2010kj} up to fully non-linear level \cite{Hassan:2011hr,Hassan:2011ea}, 
	and it relies on carefully chosen interactions 
	which are ghost-free by construction and implement the 
	so-called Vainshtein screening~\cite{Vainshtein:1972sx}. 
	Once active, this mechanism guarantees not only the recovery 
	of the GR limit, but also improves the stability of the theory under quantum 
	corrections (see, for example, Refs.~\cite{Nicolis:2004qq,Brouzakis:2013lla,deRham:2012ew,deRham:2013qqa,deRham:2014wfa}). For reviews on massive gravity see
	Refs.~\cite{Hinterbichler:2011tt,deRham:2014zqa}.

	There has been extensive work exploring cosmological solutions
	in massive gravity~\cite{Koyama:2011yg,Chamseddine:2011bu,Gumrukcuoglu:2011zh,Comelli:2012db,Gratia:2012wt,Kobayashi:2012fz,DeFelice:2012mx,Volkov:2012zb,Gumrukcuoglu:2012aa,Gumrukcuoglu:2012wt,Tasinato:2012ze,Wyman:2012iw,DeFelice:2013awa,Tasinato:2013rza,Khosravi:2013axa,D'Amico:2011jj,Solomon:2014dua} (and its bigravity generalization).
	If the graviton mass is responsible for the late time acceleration, it
	appears sensible to set it to order of the 
	Hubble parameter today or even smaller.  
	While an observationally motivated choice, there are arguments~\cite{deRham:2012ew,deRham:2013qqa} 
	as to why a small mass is a technically natural one. 
	Most stringent constraints 
	on the graviton's mass arise from the physics of solar system scales, and therefore  late time observations.

	 In the case of massive gravity, one may wonder whether 
	there exist further bounds on the 
	graviton mass from early time cosmology.  On the other hand, in bigravity, which one can think of as massive gravity equipped with an additional Einstein-Hilbert piece for the reference metric, the dynamics is richer and strongly dependent on the role played by massless as well as massive tensor modes.
	Which observables could be more sensitive to the massive (bi)gravity dynamics?
Massive (bi)gravity comes with five(seven) degrees of freedom(d.o.f.) and,
	in the current data-driven era, 
  one might want to test the presence of these additional d.o.f.s at lowest order in perturbation theory. However, the dynamics of the scalar and vector sectors can and sometimes is efficiently Vainshtein-screened, rendering the cosmological evolution almost unaltered for 
  most part of cosmic history. 
  This is precisely what one would want, but ought to verify, if any massive (bi)gravity theory is to be employed to describe late time acceleration.

	On the other hand, measuring the 
	amplitude of primordial Gravitational Waves (GWs) 	can be used to constrain various massive (bi)gravity models.
	This amplitude is traditionally parametrised 
	by the tensor-to-scalar ratio, $r$, 
	which is defined as 
	the ratio of amplitude of tensor to scalar fluctuations.
	Naively, a massive mode decays after crossing the 
	horizon during inflation, lowering the tensor power compared to
	the usual GR prediction. However, there are interesting
	subtleties related to identifying such a signal 
	in the data~\cite{Dubovsky:2009xk}. 
	In bigravity in particular, the presence of the additional tensor modes is not screened in the Vainshtein sense, calling for an investigation of this additional sector.

	In  
	massive gravity
	there are 
	naturally two metrics (for a general approach see \cite{Bernard:2015mkk}).
	If only one of the metrics is dynamical 
	then there is a clear cut massive eigenstate for the tensors and
	$r$ is sensitive to the corresponding graviton effective mass.
	In bigravity, however, both metrics  
	are dynamical~\cite{Hassan:2011zd,Hinterbichler:2012cn}, 
	resulting in two sets of coupled tensor modes with time-dependent mass eigenstates. These
	will
	contribute to the tensor
	power spectrum in a non-trivial way. It follows that the imprint of bigravity 
	on primordial GWs is necessarily 
	less transparent, 
	which has motivated further work in bigravity phenomenology in the early universe 
	\cite{Lagos:2014lca, Amendola:2015tua,Johnson:2015tfa} 
		and other contexts~\cite{vonStrauss:2011mq,Volkov:2011an,Volkov:2012cf,Berg:2012kn,Akrami:2012vf,Sakakihara:2012iq,Fasiello:2012rw,Maeda:2013bha,Volkov:2013roa,Fasiello:2013woa,Enander:2015vja}.

	We focus on a regime of the bigravity theory 
	where several stability requirements are 
	met---see Refs.~\cite{Volkov:2011an,Comelli:2011zm,Comelli:2012db,Fasiello:2013woa}, where the background cosmology was extensively studied. 
	In agreement with the perturbation analysis 
	discussed in Refs.~\cite{DeFelice:2013nba,DeFelice:2014nja},
		in that regime the Higuchi bound~\cite{Higuchi:1986py,Fasiello:2012rw,Fasiello:2013woa} is satisfied and the gradient instability \cite{Comelli:2012db} is pushed outside the reach of the effective theory \cite{DeFelice:2014nja}.

	\para{Outline}This note is organised as follows. In \S\ref{sec:review} we 
	review massive bigravity in general and specifically the conditions describing the regime of Ref.~\cite{DeFelice:2014nja}, focussing on 
	a theoretically viable region in 
	parameters space. 
	We choose consistent couplings
	to matter which fix the way gravity communicates with photons
	that go on to form the Cosmic Microwave Background Radiation (CMBR).	In \S\ref{sec:power} we  
	compute the tensor power spectrum 
	and discuss how a measurement of $r$ would
	impose constraints on massive gravity and bigravity. 
	In \S\ref{sec:signatures} we 
	elaborate on the possible $B$-modes imprints of massive bigravity in the CMBR signal.
	We summarize our work in \S\ref{sec:conclusion}. 
		
	\section{Review of massive bigravity}
	\label{sec:review}
	
	The ghost-free, non-linear extension of the Fierz--Pauli~\cite{Fierz:1939ix} 
	mass term 
	can be written
	\ba
\label{eq:LagMGR}
\L_{\rm mGR}=\frac{\mpl^2}{2} \sqrt{-g}\( R[g]+2\,m^2\sum_{n=0}^4\dfrac{\alpha_n}{n!\, (4-n)!}\  \mathcal{L}_n[\K]\)\,,
\ea
where $\mpl$ refers to the Planck mass associated with
the metric $g_{\mu\nu}$, 
$\alpha_n$ are constant interaction coefficients 
 and the composite tensors $\K$ are defined by
\ba
\K\mupn[g,f] =\delta\mupn -X\mupn\quad\quad{\rm with}\quad\quad X\mupn \equiv \(\sqrt{g^{-1}f}\)\mupn \,.
\ea
The scalar interaction potential 
is routinely and symbolically given in terms of the 
Levi--Civita tensors,
$\mathcal{L}_n[\mathcal{K}] \equiv \mathcal{E} \  \mathcal{E} \ \K^n$ where the 
summation of indices is implicit--- the interactions are
built out of characteristic polynomials of the eigenvalues of 
$\mathcal{K}$, which
is at the core of the ghost-free nature of the theory.
This theory goes generally by the name of 
\emph{massive gravity}. 

If the reference metric, $f_{\mu\nu}$, becomes dynamical, then 
the theory is referred to as \emph{bigravity}. In this case, the Lagrangian in 
Eq.~\eqref{eq:LagMGR} needs to be augmented by the kinetic term of the 
metric $f_{\mu\nu}$, 
given by the corresponding Einstein--Hilbert term (with respective Planck mass $M_f$),
which is, to date, the only known ghost-free, Lorentz-invariant kinetic term 
allowed in four dimensions~\cite{deRham:2013tfa}. Therefore, 
in bigravity models, the Lagrangian becomes\footnote{To this one generally adds minimally-coupled matter sector(s).}:
\begin{equation}
\mathcal{L}_{\rm mBiGrav}=\L_{\rm mGR} +
\frac{M_f^2}{2} \sqrt{-f}\, R[f] \ .
\label{eq:LagBiG}
\end{equation}

It is important to emphasise the following points about generic theories
of massive bigravity. 
First, notice that in these theories there is only one value for the 
bare graviton mass, $m$, which enters in the interaction potential, $\mathcal{L}$.
The two sets of helicity-2 modes generated by the two metrics $g_{\mu\nu}$
and $f_{\mu\nu}$ are associated with a massive and a massless spin 2-fields.  
The massive set of modes has an effective mass, $m_{\rm eff}$, which is
in turn related  to $m$ through the $\alpha_n$ coefficients of the mixing interactions $\mathcal{L}_n$. This $m_{\rm eff}$ is the value which enters e.g. the predictions
for observables such as 
the tensor power spectrum and 
$r$. Observational constraints 
are often directly sensitive only to ``dressed" versions of the graviton mass, a fact we shall make use of repeatedly in this note.

We also stress that bigravity theories describe two dynamical metrics while being entirely consistent with Weinberg's theorem~\cite{Weinberg:1964ew,Weinberg:1965rz} which argues that no two massless spin-2 fields mediating the long-force gravitational interaction can coexist. In bigravity there is a massive and a massless field, 
so that no violation of such theorem occurs.

\subsection{Bigravity cosmology}

The task of finding viable massive (bi)gravity cosmological solutions has recently received considerable attention, generating a rich and interesting literature~\cite{Lagos:2014lca, Amendola:2015tua,Johnson:2015tfa} 
		and other contexts~\cite{vonStrauss:2011mq,Volkov:2011an,Volkov:2012cf,Berg:2012kn,Akrami:2012vf,Sakakihara:2012iq,Fasiello:2012rw,Maeda:2013bha,Volkov:2013roa,Fasiello:2013woa}.
A no-go theorem exists for FLRW solutions in massive gravity \cite{D'Amico:2011jj}: there are no FLRW solutions in a massive gravity theory with a Minkowski reference metric, $f_{\mu\nu}=\eta_{\mu\nu}$. Far from being a problem, this realization has lead to further work in several different directions.

Within massive gravity, one may relax the exact homogeneity or isotropy assumption by introducing some degree of inhomogeneities in the \stu fields \cite{D'Amico:2011jj}, consider open (closed) FLRW solutions~\cite{Gumrukcuoglu:2011ew,Gumrukcuoglu:2011zh,Vakili:2012tm,DeFelice:2012mx}, investigate the cosmology arising from a different choice for $f$ \cite{Fasiello:2012rw,Langlois:2012hk}  or study an extended, ghost-free version of the theory preserving the same five degrees of freedom \cite{deRham:2014gla}. 
Naturally, another way out of the no-go is that of considering theories with additional degrees of freedom. Massive bigravity and generic multi-metric theories, the quasi-dilaton model \cite{D'Amico:2012zv,DeFelice:2013tsa,Kahniashvili:2014wua,Motohashi:2014una,Heisenberg:2015voa} all belong to this class. Alternatively, 
one may choose a non-trivial coupling to matter~\cite{deRham:2014naa}.

In any of the above contexts, once the background solutions have been found, a lot of care must be exerted before declaring those solutions viable. Setting aside for a moment observational constraints, the study of background and perturbations alone might indeed reveal some pathologies. A healthy branch of solutions may support vanishing kinetic terms (the origin of strong-coupling issues) or it may lead to perturbations which do not satisfy the unitarity (Higuchi) constraint or present a gradient instability.

Complementing these requirements are those stemming from observational viability: in short one would want the ``background'' cosmic evolution to be essentially ascribable to 
GR up to the latest era, that of dark energy domination. One should add that, in order to pass, for example, solar system tests, an active and effective Vainshtein mechanism also needs to be in place. 

In what follows 
we will review the analysis and employ the notation of Refs.~\cite{DeFelice:2013nba,DeFelice:2014nja}. Our interest lies in the flat, homogeneous and isotropic FLRW universe.  We take
\ba
\label{eq:frw}
\d s_g^2 = g\mn \d x^\mu \d x^\nu = -N^2(t) \d t^2+a^2(t)\ \d \vec{x}^2\\
\d s_f^2 = f\mn \d x^\mu \d x^\nu = -\mathcal{N}^2(t) \d t^2+b^2(t)\ \d \vec{x}^2\,,
\ea
respectively for each metric. The background Friedmann equations read
\begin{eqnarray}
&& 3 H_g^2  = 
m^2\,\rhomg +\frac{\rho_g}{M_g^2}\,,
\label{eq:Friedg}\\
&&3 H_f^2  =
\frac{m^2}{\kappa}\,
\rhomf +
\frac{\rho_f}{\kappa\, M_g^2}
\,,
\label{eq:Friedf}
\end{eqnarray}
where $H_g\equiv \dot{a}/aN$ and $H_f\equiv \dot{b}/b \mathcal{N}$
are the Hubble parameters, $\kappa \equiv M_f^2 / \mpl^2$, 
and $\rho_g$ and $\rho_f$ are the energy density associated 
with matter coupling directly to the metric 
$g_{\mu\nu}$ and $f_{\mu\nu}$, respectively. It will be convenient for simplicity
 of notation to define the following dimensionless quantities 
\begin{equation}
 \rhomg  \equiv   
  U(\xi)-{\xi\over 4}\ U'(\xi)
\,,
  \qquad
 \rhomf  
\equiv
{1 \over 4\xi^3}\  U'(\xi)\, , 
\qquad \hat{\rho}_m\equiv  \rhomg(\xi)- \frac{\xi^2}{\kappa}\rhomf(\xi)
\ ,
\label{bgdefs}
\end{equation}
with primed variables being differentiated with respect to 
$\xi= \xi(t) \equiv b(t)/a(t)$, and 
\begin{equation}
 U(\xi)\equiv -\alpha_0+4(\xi-1)\alpha_1-6(\xi-1)^2\alpha_2+4(\xi-1)^3\alpha_3-
(\xi-1)^4\alpha_4\, ,
\label{Udef}
\end{equation}
where $\alpha_n$'s correspond to the interaction coefficients, 
cf. Eq.~\eqref{eq:LagMGR}. To avoid a number of pathologies 
described in Refs.~\cite{Comelli:2012db,Comelli:2014bqa,DeFelice:2014nja}, we choose the so-called \emph{healthy branch}\footnote{See e.g. \cite{Comelli:2014bqa,Gumrukcuoglu:2011zh,DeFelice:2012mx} for a more detailed analysis on choosing branches.} of cosmological
solutions which corresponds to the dynamical equation
\begin{equation}
H_g = \xi \ H_f\ .
\label{bianchi}
\end{equation}
It will also be convenient for our subsequent analysis to define
\ba
 J(\xi)  \equiv  
{1 \over 3}\left(U(\xi)-{\xi\over 4}\ U'(\xi)\right)' 
\ .
\label{Jdef}
\ea
It follows from this definition that $ J(\xi)$ is an implicit function of the $\alpha_n$ coefficients which 
set the strength of the several interactions in the Lagrangian~\eqref{eq:LagMGR}.

	\subsection{The low-energy regime as a viable region in parameter space}
	\label{subsec:limit}
In bigravity, 
the mass eigenstates corresponding to the massive modes are generically time dependent~\cite{Fasiello:2013woa}. In order to isolate the massless and massive tensor modes at each given time one ought to diagonalise the modes using an appropriate basis. It is a convenient feature of the region of the parameter space studied in Ref.~\cite{DeFelice:2014nja} that the effective mass in this regime is actually constant, which makes the analytic derivation of observable quantities, such as $r$, much more tractable.

We shall start by defining the low-energy limit regime (henceforth LEL), investigated in Ref.~\cite{DeFelice:2014nja}, as that satisfying the following  condition 	
	\begin{equation} 
\frac{\rho_g}{m^2 \, M_g^2} \ll 1\quad {\rm and} \quad
 \frac{\xi^2\, \rho_f}{\kappa \, m^2 \, M_g^2} \ll 1\,.
\label{eq:lowenergy}
\end{equation}
Implementing both these relations as well as Eqs.~\eqref{eq:Friedg} and \eqref{eq:Friedf} is of great consequence for the value of the bare mass $m$ and the role of the interactions in $\rhomg$. The realisation that 
GR provides an accurately verified description for most of cosmic history predating the dark energy era demands that $\rho_g$ dominates the RHS of the Friedmann equation correspondingly. This, together with the regime in Eq.~\eqref{eq:lowenergy} 
translates into the following conditions:
\ba
m^2\gg H_g^2  \, , \qquad    \rhomg < H_g^2/m^2\, \qquad  \Rightarrow\qquad  \rhomg \ll 1
\label{b1}
\ea

Some of the freedom that comes with having the $\alpha_n$ coefficients in  $\rhomg$ is then being put to use in taming the ``large" value of the bare mass $m$ in Eq.~(\ref{eq:Friedg}). We continue below navigating the parameter space of the LEL regime to expose its dynamical consistency and, at the same time, underline the bounds the latter demands.

Combining Einstein equations with Eq.~(\ref{bianchi}) one derives the following:
\ba
\hat{\rho}_m(\xi) =-\frac{\rho_g}{m^2 M_g^2}+\frac{\xi^2 \rho_f}{\kappa m^2M_g^2}\quad \Rightarrow \quad  \hat{\rho}_m(\xi)\Big|_{LEL} \ll 1\, .
\label{rhohat}
\ea
It is intuitively clear then that in the LEL a solution to Eq.~(\ref{rhohat}), seen as a dynamical equation for $\xi$, is that of a constant $\xi=\xi_c$ satisfying $\hat{\rho}_m(\xi_c)=0$.\footnote{If the solution in the low energy regime is not such that $\hat{\rho}_m(\xi_c)=0$, then $\hat{\rho}_m(\xi_c)$  can always be reabsorbed into a cosmological constant contribution.}
 As to the value of $\xi_c$,  stability considerations we will touch upon later on suggest it be order unity. 
To fully specify the LEL regime, one needs to further require the following relation holds true:
\ba
1\gg \Bigg| \frac{\kappa \xi_c \rhomg}{J(\xi_c)} \Bigg| \sim (\kappa\xi_c) \,\frac{\rhomg(\xi_c)}{\rhomg^{\,\prime}(\xi_c)}
\label{b2}
\ea
This can be interpreted as follows: it corresponds to requiring $|m^2 \hat{\rho}_{mg}/m^2_{\rm eff}|\ll1$. It holds in view of the low-energy Higuchi bound,  $(m_{\rm eff}^2 \gtrsim 2H^2)$, and Eq.~(\ref{eq:Friedg}) and can hold in eras preceding dark energy domination.
Condition~(\ref{b2}) may be arrived at by enforcing $\kappa\ll 1$ or by appropriately choosing the $\alpha_n$ coefficients  within $\rhomg$ to enforce $\rhomg \ll \rhomg^{\prime}$. 
Note that the latter option is not light on the $\alpha_n$s because of the requirement already in place via Eq.~(\ref{b1}).

We note in passing that the lack of gradient instability 
in the scalar sector corresponds to   
implementing
\ba
c_s^2\sim 1+ \frac{2(\tilde{c}-1)}{3}\frac{d\,ln\,J}{d\,ln\, \xi}-\frac{2 \xi^2 (\rho_f + P_g)}{3 \kappa M_g^2 m_{{\rm eff}}^2 } > 0 \, , \quad {\rm with}\quad  m^2_{{\rm eff}}\equiv m^2 \ \frac{1+ \kappa  \xi^2 }{ \kappa \xi^2 } \Gamma(\xi)\ ,
\label{eq:defmeff}
\ea 
and where 
\ba
 \Gamma(\xi) \equiv \xi \, J(\xi) + \frac{(\tilde{c} -1) \xi^2}{2} \, J'(\xi)\ .
 \label{eq:defGamma}
\ea
The unitarity bound is automatically satisfied\footnote{See Refs.~\cite{Fasiello:2012rw,Fasiello:2013woa,Higuchi:1986py,Yamashita:2014cra} for more details.} upon enforcing Eqs.~\eqref{b1} and \eqref{b2}. Note the appearance of the effective mass, $m_{{\rm eff}}$, which is the physically relevant quantity here and corresponds to the bare mass dressed by the non-linear interactions which make up the $\mathcal{U}_n$ potentials. For later use we point out here that already at an intuitive level it is clear that a large effective mass would give way to a dynamics dominated by the massless modes. One may borrow the familiar -integrating out d.o.f.- physical picture to guide this intuition \footnote{See also \cite{Akrami:2015qga} on this point.}.

As we have seen, working in the LEL regime comes with a simplified dynamics. It allows for example an expansion of all relevant quantities about $\xi=\xi_c$, as a result of which one  
can write the Friedmann equation in Eq.~\eqref{eq:Friedg} as
\begin{equation}
3 H_g^2\simeq \dfrac{\rho_g }{\mpl^2 (1+\kappa \, \xi_c^2)}
+\Lambda_{\rm eff}\ ,
\label{eq:fried2}
\quad {\rm where} 
\quad \Lambda_{\rm eff}\equiv \Lambda \(
1+\dfrac{2\kappa \,\xi_c \,(\xi-\xi_c)}{1+\kappa \xi_c^2}
\) \, ,
\end{equation}
and with $\Lambda \equiv m^2\rhomg (\xi_c)$. 
We can thus write
	\begin{equation}
3 H_g^2\simeq \dfrac{\rho_g}{\mpl^2 (1+\kappa \, \xi_c^2)}\ ,
\label{eq:fried3}
\end{equation}
a relation we will put to use in \S\S \ref{sec:power} and \ref{sec:signatures}.	
	 In obtaining our estimates for the tensor-to-scalar power spectrum, we require the LEL regime to be a good approximation at the very least as early as the onset of recombination, when the imprint of bigravity signatures on the CMB occurs. It is easy to verify this scale is well within even the most cautious estimate for the strong coupling scale $\Lambda_{3,{\rm f}}=(m^2 \kappa^{1/2}M_g)^{1/3}$.
	
		\subsection{Coupling massive (bi)gravity to matter}
    \label{sec:matter}
    Before we compute observables, we need to specify 
    the coupling of gravity to matter.
    When the universe 
    was 380,000 years-old, recombination occurred causing 
    the photons to decouple from the hot plasma 
    and free-stream up to the present time. 
    The photons of this era are the CMBR we observe 
    in the microwave sky 
    and they 
    contain a snapshot of the physics of the early universe. 
    They may then
    encode information about a theory of gravity which is other than GR.

    How does massive gravity
    couple to the matter sector?
    Classically, it 
    was shown by Hassan \& Rosen~\cite{Hassan:2011zd}
    that bigravity can couple covariantly to matter 
    without reintroducing a ghostly degree of freedom.
       At the quantum level, one may worry that the irrelevant 
    interactions which make up the non-linear theory might not be 
    under control.\footnote{In general, the quantum stability of field 
    theories which rely on large derivative self-interactions, 
      as is the case of massive gravity, is 
    not trivial. However, recent work has shed some light on  
   the role of the quantum mechanical realisation of 
   the Vainshtein mechanism in protecting the theory---see Refs.~\cite{Nicolis:2008in,Nicolis:2004qq,dePaulaNetto:2012hm,Brouzakis:2013lla,Brouzakis:2014bwa,deRham:2014wfa}.}
    
    In bigravity or multigravity theories, 
    one could wonder whether all the 
    dynamical metrics could democratically and covariantly couple to the \emph{same} 
    matter sector.
    Unfortunately this induces
    a ghost at an 
    unacceptably low scale~\cite{Yamashita:2014fga,deRham:2014naa}, 
   making the model
    unstable. As a result, 
    either each dynamical metric couples to its own matter sector, or 
    the coupling of gravity to matter is made through a composite
    metric. In the last case, however, 
      FLRW ansatzs for both $g_{\mu\nu}$ and $f_{\mu\nu}$ 
    require matter to which they couple to be identified as a dark sector, \cite{deRham:2014naa,deRham:2014fha}
       which can later on couple to standard matter. This would require knowledge of this additional coupling.
    
      For simplicity, in what follows we couple matter to one of the metrics, $g_{\mu\nu}$. 

   The matter Lagrangian is fully described by
     \begin{equation}
    S_{\rm matter}=\int{\d^4 x \  \sqrt{-g} \ \mathcal{L}_{\rm matter} (g, \phi_i)}\ ,
    \label{eq:matteraction}
    \end{equation}
where $\phi_i$ labels the matter field species, including 
photons which encode the physics of the early universe after recombination occurs.
	
    \section{Primordial gravitational waves}
    \label{sec:power}
    
    Primordial gravitational waves are seeded by tensor modes of the
    primordial perturbation. In massive gravity, there is only one family of modes, 
    which we shall denote by $h_{\mu\nu}$ whereas for bigravity there are two sets 
    of modes at play, 
    $h_{\mu\nu}$ and $\ell_{\mu\nu}$. Therefore,
    the fluctuations of both metrics, $g_{\mu\nu}$ and $f_{\mu\nu}$, now
    contribute to the power spectrum of primordial GWs.

    We start by reviewing the cosmological perturbation 
    analysis performed by De Felice et al.~\cite{DeFelice:2014nja}.
    We write the metric in ADM variables
    \ba
    g_{\mu\nu}=-N^2 \d t^2+a(t)^2 (\gamma_{ij}+h_{ij}) \d x^i \d x^j
    \quad \textrm{and}
    \quad
    f_{ij}=-\mathcal{N}^2 \d t^2+ b(t)^2 (\gamma_{ij}+\ell_{ij}) \d x^i \d x^j
    \ea
    where $\gamma_{ij}$ is spatially flat, $a$ and $b$ are 
    scale factors and 
    $N$ and $\mathcal{N}$ are lapse functions 
    corresponding to each space-time metric.
    The quadratic action of the tensor modes becomes 
        \ba
S^{(2)}_{\rm tensors} &= \dfrac{\mpl^2}{8} \displaystyle{\int \d^4 x} \,N a^3\sqrt{\gamma} \Bigg[ &
\frac{\dot{h}^{ij}\dot{h}_{ij}}{N^2}+\frac{h^{ij}}{a^2}\nabla^2 \ h_{ij}
+
\kappa\, \tilde{c}\, \xi^4
\left(\frac{\dot{\ell}^{ij}\dot{\ell}_{ij}}{\mathcal{N}^2} +\frac{\ell^{ij}}{b^2}
 \nabla^2 \ \ell_{ij}\right) \nn \\
&&- m^2\, \Gamma(\xi) 
\left(h^{ij}-\ell^{ij}\right)\left(h_{ij}-\ell_{ij}\right)
\Bigg]\, ,
\label{eq:S2tensors}
\ea
where $\gamma$ can be set to unity, $\nabla^2$ denotes the spatial Laplacian 
and 
\begin{equation}
 \tilde{c}  \equiv \frac{\mathcal{N} a}{N b}\ .
\end{equation}
Dotted quantities are differentiated with respect to cosmic time.
The action above~\eqref{eq:S2tensors} features non-trivial 
interactions between $h$ and $\ell$ fluctuations, which makes 
analytical predictions far from straightforward. To make progress, it is
be useful to find a regime in which a diagonalisation of this action
is possible, as argued before.

     Considering only the transverse and traceless perturbations as the 
    relevant degrees of freedom, it is convenient to choose a different basis
    for the tensor perturbations as follows \cite{DeFelice:2014nja}:
   \begin{equation}
      H^-_{ij} \equiv h_{ij}-\ell_{ij} \quad 
      {\rm and}
      \quad
      H^+_{ij} \equiv 
\frac{h_{ij}+\kappa \, \xi^2 \ell_{ij}}{1+\kappa \, \xi^2}\ .
\label{eq:diagbasis}
    \end{equation} 
We stress that the validity of this diagonalization is limited to the LEL regime. The mass eigenstates of a bigravity theory are in general time dependent and so will be the diagonalization basis. Nevertheless, in what follows we will employ the customary inflationary normalization for the tensor wave functions whose behaviour at recombination is what determines possible CMBR imprints.

Before proceeding with the analysis of tensor modes in the LEL configuration we pause here to stress that any instability in the tensor sector should not go unnoticed and it is different in nature from the one concerning the other d.o.f.s. Indeed, no matter how efficient the screening of scalars and vectors, the two copies of the tensors at hand can be safely considered unscreened.
The act of normalizing things ``as usual" for inflation and considering the tensors diagonalized from the onset of the LEL may be insensitive to part of the dynamics which goes beyond the LEL regime\footnote{This is important because the coupling among the tensor modes, although very small, in time can have significant effect once higher order in perturbations are taken into account. This again points to the importance of the initial value problem as  emphasized in \cite{Lagos:2014lca}. See also \cite{Johnson:2015tfa} for a step in this direction. Note that the scales probed by \cite{Johnson:2015tfa} are far beyond the reach of the LEL regime.}, as is exemplified by the work in \cite{Lagos:2014lca, Cusin:2014psa}, which respectively focused on different branches of solutions.

Working at lowest order in perturbations, the simplified action for the tensor modes becomes 
\ba
S^{(2)}_{\rm tensors} & = & 
\frac{1}{8}
\int \d^4 x \, a^3\,\Bigg[
M_+^2 \left(
\dot H_+^{ij}\dot H^+_{ij}
+\frac{H_+^{ij}}{a^2} \, \nabla^2 H^+_{ij}
\right)
\nonumber\\
&&\qquad\qquad\qquad \qquad\qquad 
+M_-^2\left(
\dot H_-^{ij}\dot H^-_{ij}
+\frac{H_-^{ij}}{a^2} \,\nabla^2\, H^-_{ij}
- m_{\rm eff}^2 
H_-^{ij} H^-_{ij}
\right)\Bigg]\, .
\label{eq:diagonalbasis}
 \ea   
 As we shall see shortly, it is this mass parameter $m_{\rm eff}$
 associated with the tensor modes which
 enters observable quantities and which can be used to constrain
 the parameters of the theory.
 The normalisation of the kinetic terms 
 is set by the mass scales $M_{+}$ and $M_{-}$ given by
\ba
M_{+}^2\equiv (1+\kappa \,\xi_c^2)\, \mpl^2 
\quad
\textrm{and}
\quad
M_{-}^2\equiv \dfrac{\kappa \, \xi_c^2}{(1+\kappa\, \xi_c^2)^2} \ M_{+}^2\ .
\label{eq:defMplusMminus}
\ea 

In perturbation theory
the matter Lagrangian~\eqref{eq:matteraction} is given, to lowest order, by
    \begin{equation}
    \delta S_{\rm matter}=\int{\d^4 x \  h_{\mu\nu} T^{\mu\nu}}\ .
    \label{eq:mattercoupling}
    \end{equation}
The usefulness of the $\left\{+,-\right\}$ basis is clear: $H_{ij}^+$ is the massless mode
whereas the massive mode is $H_{ij}^{-}$. 
In light of our choice of coupling to matter~\eqref{eq:mattercoupling}, 
photons do not couple to a massive nor massless graviton. 
Instead, they couple to a linear combination of massive and massless 
modes in the diagonal basis described in Eq.~\eqref{eq:diagonalbasis}.
Importantly, the massive mode is associated with a redressed mass
$m_{\rm eff}$, which albeit related, is \emph{not} the graviton mass. 

    In terms of the diagonalised variables of 
    Eq.~\eqref{eq:diagbasis}, the coupling to the matter sector is
    \begin{equation}
       \delta S_{\rm matter}=  \int{\d^4 x \
       \left\{
      H_{ij}^{+} + \dfrac{\kappa\,\xi_c^2}{1+\kappa \, \xi_c^2} H_{ij}^{-}
       \right\} T^{ij}}\ .
    \label{eq:mattercouplingdiag}
    \end{equation}
This means that gravity and the history of the universe will, 
in principle, change. To 
guarantee that the cosmology remains invariant so as 
to agree with $\Lambda$CDM and to reproduce the
Newtonian limit, we need to 
require that the 
Planck mass
(or equivalently the Newton's gravitational constant) 
is essentially the same as 
measured in solar system scales. Consider two mass tests 
described by the energy-momentum tensor above and subject to gravity set 
by this matter coupling. One may easily derive that 
the effective Planck mass associated with this 
modified gravitational force is actually the Planck mass
associated with the metric $g_{\mu\nu}$:
\ba
M_{\rm Pl, eff}\equiv \mpl \ .
\ea

So far, all we have assumed within the LEL was that $\xi_{c}\sim\Or(1)$, 
but there were no stringent constraints on the value allowed for $\kappa$. 
If we demand 
the cosmological evolution to be the same for most of cosmic history, as well as negligible modifications to the 
Newton's gravitational constant, then from
the Friedmann equation~\eqref{eq:fried3}, we ought to require
\ba
\kappa\ll \xi_{c}^{-2}\sim \Or(1)\ .
\label{kappa}
\ea

Considering the uncertainty associated with the empirical determination
of the value of the Newton's gravitational constant, 
we estimate $\kappa$ to be $1$ part in $100,000$.  One might be worried by the realization that this regime corresponds to a small $M_f\ll M_{Pl}$.  This results, in turn, in the lowering of the naive strong coupling
scale $(m^2 M_f)^{1/3}\equiv \Lambda_{3,f}$.\footnote{Here we are focusing on the case when the strong coupling scale can be as low as $(m^2 M_f)^{1/3}$. Being a very low scale, this could prove difficult for the phenomenology at and above that energy scale.} 
This worry is legitimate and we just note here that the scale we are negotiating with is that of $H_{r}$. One can easily check that even if the strong coupling scale is as low as 
$\Lambda_{3,f}$, the theory is still predictive at the time of recombination, 
for which the graviton mass can be chosen such that $H_{\rm r}\ll\Lambda_{3,f}$, a condition which is automatic in the LEL regime.

    \para{Other d.o.f}  Before proceeding with the study of observables related to tensor perturbations we pause here to comment on the role played by the other degrees of freedom in our setup. Clearly massive gravity alone already spans 5 d.o.f. whose effect on cosmological evolution up to recombination must be accounted for. So far we have largely neglected the dynamics of the scalar (helicity-0) and vector (helicity-1) sector. Their dynamics has already been the subject of careful investigations at the lowest orders in perturbation theory hinting at early(late) time instabilities. However, such an analysis may or may not capture the full physical picture, especially if limited to low orders in perturbation theory. It has further been suggested that a way around these issues may be found in a restricted pool of favourable initial conditions and that the so-called initial value problem for bigravity needs further study  \cite{Lagos:2014lca} (see also Ref.~\cite{Brito:2014ifa} for related work).   We can only but agree on this latter point. A step in this direction was taken in \cite{Johnson:2015tfa} where initial conditions are discussed in an inflationary context. This implicitly assumes that the effective strong coupling scale is as high as $H_{\rm inf}$.

 In this manuscript we take the view that, as is often the case in solar system dynamics, an efficient\footnote{Incidentally, we note that an active screening in the LEL demands \cite{DeFelice:2013nba} that $|\d{\rm ln}J(\xi)/\d{\rm ln} (\xi)|\gg 1$, see Table 1.} Vainshtein mechanism will make use of the non-linearities of the theory to milden the role played by the scalar(vector) sector in the cosmological evolution, effectively screening them. Although suggestive results are present in the literature, at this level this is indeed just an assumption we make here, and not a small one at that.  
 
We now turn to the observable spectrum arising
    in this bigravity model. 
    The coupling in Eq.~\eqref{eq:mattercouplingdiag} 
    dictates the power transmitted by the tensor modes (both from the 
    massive and the massless spin-2 fields)
    to the CMBR photons. Before presenting the formula for 
    $r$, we make a short digression into the 
    power spectra of massive and massless tensor modes which would individually
    be imprinted in the CMBR. For bigravity theories, the power spectrum will be 
    a composite measure of both signals. 
    For the purposes of our estimates, it suffices to work
    at lowest order in the slow-roll approximation, which is assumed to hold throughout 
    inflation.

    \para{Power spectrum of massless modes}The analysis of the perturbation theory for massless tensor modes is well known
    in the literature~\cite{Copeland:1993jj}. 
    In bigravity theories, the massless mode is represented by 
    $H^{+}$ in Eq.~\eqref{eq:diagonalbasis} and 
    their evolution is fixed by the time
    they cross the horizon, after which their amplitude remains constant 
    thereafter. It is therefore sufficient to determine their 
    super-horizon evolution. On super-horizon scales, the resulting power spectrum is given by
    \ba
    P_{\rm massless}\equiv 
    \langle  H_{ij}^{+} (\vec{k}_1)\, H_{jl}^{+} (\vec{k}_2)\rangle
     \equiv \dfrac{P_{\rm t, GR}}{1+\kappa \xi_c^2}.
   \label{eq:Pmassless}
    \ea
    Here we adopt the usual notation that starred quantities are 
    evaluated at the time of horizon crossing for the mode $k$, and 
    we use $H\equiv H_g$ to avoid clutter.
    Notice the different normalisation arising from Eq.\eqref{eq:diagonalbasis} of the tensor spectrum 
    from the GR one, which we have denoted by $P_{\rm T, GR}$, where 
    the only scale is $H$.

     \para{Power spectrum of massive modes}Massive tensor modes 
     have their evolution determined by their corresponding mass horizon, and they do not
     get frozen on super-horizon scales but rather oscillate. These modes are denoted
     by 
    $H^{-}$ in Eq.~\eqref{eq:diagonalbasis} and give the following power spectrum 
     \ba
      P_{\rm massive}\equiv 
      \langle  H_{ij}^{-} (\vec{k}_1)\, H_{jl}^{-} (\vec{k}_2)\rangle
    \equiv \dfrac{1+\kappa \, \xi_c^2}{\kappa \xi_c^2}\, P_{\rm t,mGR} \,
     \label{eq:Pmassive}
     \ea
     where $P_{\rm t,mGR}$ stands for the power spectrum obtained from 
     the standard GR action for tensors 
     where there is an additional mass term (in this case $m_{\rm eff}$).
     Notice again the different non-trivial normalisation of the power spectrum (arising from Eq.\eqref{eq:diagonalbasis}).

\subsection{Tensor-to-scalar ratio in bigravity}The power transmitted by the tensor modes
in this class of theories is governed by how they couple to matter via 
Eq.~\eqref{eq:mattercouplingdiag}. It follows that the primordial power spectrum of tensors is
\ba
\langle h_{ij}(\vec{k}_1)\, h_{jl}(\vec{k}_2) \rangle =
\langle H_{ij}^{+}(\vec{k}_1)\, H_{jl}^{+}(\vec{k}_2) \rangle 
+
\left(\dfrac{\kappa \, \xi_c^2}{1+\kappa \, \xi_c^2} \right)^2 \
\langle H_{ij}^{-}(\vec{k}_1)\, H_{jl}^{-}(\vec{k}_2) \rangle 
\label{relative}
\ea
with the individual power spectra given by Eqs.~\eqref{eq:Pmassless} and \eqref{eq:Pmassive}.
The fact that this prediction is different from that of GR 
is only one source of 
modification to the tensor-to-scalar ratio, which is also sensitive to the scalar 
perturbations. Since the Friedmann equation is still slightly changed, this will reflect
on the matter power spectrum and therefore on the primordial scalar power spectrum.
In particular, from Eq.~\eqref{eq:fried3} the energy density in matter is
\ba
\rho_{\rm matter} = \dfrac{\rho_g}{1+\kappa \, \xi_c^2}\ ,
\ea
which results in the scalar power spectrum changing to
\ba
P_{\rm s, biGrav}=\dfrac{P_{s,GR}}{1+\kappa \, \xi_c^2}\ ,
\ea
where the subscript $s$ refers to scalar.
Consequently, these theories predict 
\ba
r_{\rm BiGrav}=r_{\rm GR}
+
(\kappa\,\xi_c^2) \ r_{\rm mGR}  \ ,
\label{eq:rbigrav}
\ea
where $r_{GR}$ denotes the tensor-to-scalar ratio in GR (that is, assuming a massless
graviton) whereas $r_{\rm mGR}$ corresponds to that in a theory of a massive graviton.
Under the assumption that 
the cosmology is not significantly changed
and in the limit where $\kappa \xi_c^2\ll 1$, it follows that the prediction for $r$ in these 
theories is the same as the original for GR.

\subsection{Summary on bounds in parameter space}
To obtain the observable predictions in bigravity theories, our 
starting point was the region in parameter space in the theory where 
known instabilities are absent~\cite{DeFelice:2014nja} which we have further restricted due to the $\kappa \ll1$ requirement. 
Our results reflect this choice. We summarise on Table~\ref{table:regime} the parameter space available so that the theory is free of known instabilities. This entails that, even though there is in principle more freedom to choose from the higher number of degrees of freedom in bigravity then in massive gravity, 
absence of instabilities plays a rather restrictive role.

\begin{table}[ht]
	\heavyrulewidth=.08em
	\lightrulewidth=.05em
	\cmidrulewidth=.03em
	\belowrulesep=.65ex
	\belowbottomsep=0pt
	\aboverulesep=.4ex
	\abovetopsep=0pt
	\cmidrulesep=\doublerulesep
	\cmidrulekern=.5em
	\defaultaddspace=.5em
	\renewcommand{\arraystretch}{1.8}
	\begin{center}
		\small
\rowcolors{1}{}{lightgray}
		\begin{tabular}{ccc}
			\toprule
			{\bf Stability requirement} & \qquad {\bf Parameters} &\qquad 
			{\bf Bound}
			\\
			\cmidrule{1-3}
			\cellcolor{lightgray} Higuchi bound & \cellcolor{lightgray} \quad   $m_{\rm eff}$	&	\cellcolor{lightgray} $m^2_{\rm eff} > 2H_g^2$	\\[2mm]
			\cmidrule{1-3}
			Absence of strongly coupled perturbations & \qquad $H_g$ &
			$H_g=\xi \, H_f$ \tablefootnote{This condition specifies the so-called healthy branch.}
			\\
			\cmidrule{1-3}
			\cellcolor{lightgray} No gradient instabilities \& active Vainshtein & 
		\cellcolor{lightgray} \quad   $m, H$	&	\cellcolor{lightgray} 		$m\gg H$, $|\d{\rm ln}J(\xi)/\d{\rm ln}(\xi)|\gg 1$\\[2mm]
			\cmidrule{1-3}
			Standard Friedmann evolution & \qquad $M_g$ &
			$\kappa \ll 1$ with $\xi\sim \mathrm{O}(1)$ \tablefootnote {This last assumption is to be understood within the LEL~\cite{DeFelice:2014nja}.}\\
 			\bottomrule
		\end{tabular}
	\end{center}
	\caption{Collection of stability bounds for the phenomenology discussed in this paper. Notice that these conditions correspond to relatively strict bounds on the $\alpha_n$ interaction coefficients in the massive bigravity Lagrangian. This can be easily seen by expressing $J$ in terms of $\alpha_n$ via Eqs.~(\ref{Udef}), (\ref{Jdef}).   }
	\label{table:regime}
	\end{table}

The tensor-to-scalar ratio is not the only observable which can provide 
constraints to this class of theories. Interesting imprints in the CMBR are 
also forecast which, provided they affect observable modes, can reveal 
interesting physics about the early universe. We turn next to such 
characteristic signatures in the $B$-modes signal, which were
originally unveiled in Ref.~\cite{Dubovsky:2009xk}. 
We adapt their results to discuss their implication on 
the bigravity models we discussed here.

\section{Possible Signatures in the CMBR spectrum}
\label{sec:signatures}

The study of the CMB spectrum plays a fundamental, unparalleled role in modern cosmology. It essentially represents an open window over the past of the Universe, taking one all the way back to the \textit{recombination} epoch. It is then clear that, in the study of how a massive(massless) graviton may affect the CMB radiation,  the crucial scale one will have to negotiate with is the size of the Hubble radius at recombination, $H_r^{-1}$.\footnote{The contribution from reionisation is relevant for small $\ell$. In the massless case $\ell >20$ is sufficient to ignore this contribution. The numerical analysis of \cite{Dubovsky:2009xk} supports the approximation scheme neglecting the reionization contribution also for a large range of effective mass values.}  As we shall see below, a simple description emerges for the primordial tensor sector contributions to B-modes: one can think of two main regimes for the effective tensor mass $m_{{\rm eff}}$, with the transition value for $m_{{\rm eff}}$ being $H_r$ itself.

In deriving the results for this section we will heavily rely on the work in \cite{Dubovsky:2009xk}. In there, the authors present a phenomenological approach aimed at placing bounds on the graviton mass which can be adapted to our case. Their results are quite general and hold true for the tensor sector as long as the tensor wavefunction satisfies the usual tensor e.o.m., just as the one that follows from  Eq.~(\ref{eq:diagonalbasis}). As per \textit{Section} \ref{sec:review}, the model under scrutiny here is a bigravity theory, it counts seven degrees of freedom which include massive and massless tensor modes. 

As documented in Eq.(\ref{eq:diagbasis}), both the massless and the massive tensor modes, decoupled from one another in the $^{+/-}$ basis, directly couple with matter. The crucial realization in what follows is that there exist important regions in the parameters space where, to a good approximation, the result of the two contributions is additive not just at the level of the action\footnote{This is trivially true as we linearly couple gravity to matter in Eq.~(\ref{eq:mattercoupling}).}, but it remains so up the source term $|\psi|^2$ and therefore eventually propagates all the way to the expression for the coefficients $C_{BB,l}^{T}$ . We will argue in particular that this is the case in the mass regime which would in principle generate the most prominent effect in the CMB.

As we have seen, it turns out that in the LEL, a regime we chose in order to avoid known instabilities, the relative coefficient regulating the contribution of the massive tensor modes (as opposed to the massless ones) is very small because it depends linearly on $\kappa \ll1$. This will result in a very hard-to-detect massive tensor sector. We will see below that in a specific mass range a massive graviton actually enhances the gravitational signal by two orders of magnitude as compared to the massless case. However, in our case this enhancement is no match for the $\propto \kappa$ suppression. We nevertheless provide a detailed discussion also in the hope that it will be later applied to a scenario where the relative coefficient is order one so that ``massive" imprints would be more conspicuous.

\subsection{B-modes }

\begin{center}
\begin{figure}[h]
\begin{center}
 \includegraphics[width=0.6\textwidth]{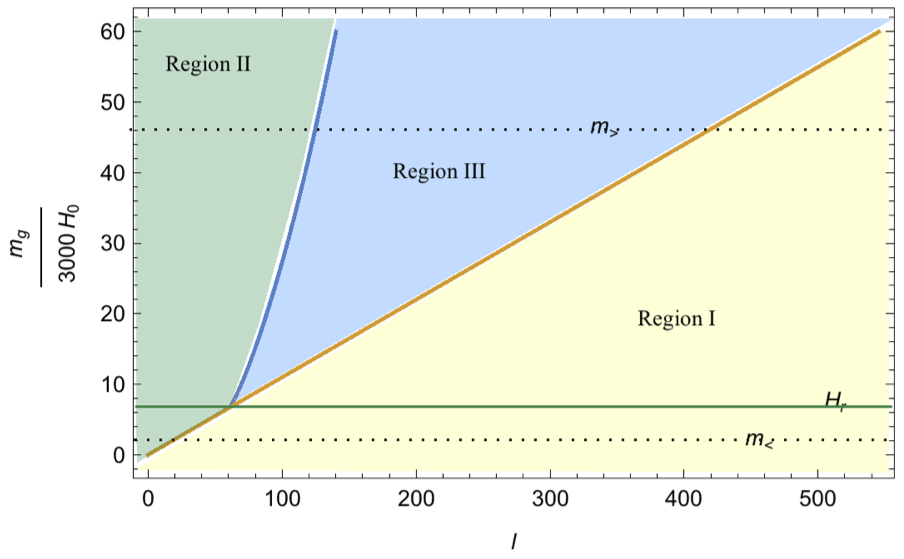}
  \caption{Above is  the effective mass $m_{{\rm eff}}$ in units of the Hubble rate today, $H_0$, versus  multipole momenta $\ell$. The upper limit of $Region$ I is obtained by  requiring that the multipole $\ell$, for which  $\ell \sim q(\tau_0-\tau_{{\rm r}})$, coincides with $\ell_0$, that is the multipole corresponding to a physical momentum the size of the mass $\meff$ at recombination. 
 The border between  $Region$ II and $Region$ III is obtained by identifying the multipole $\ell_m$ corresponding to the momentum that becomes non-relativistic at the same time it enters the horizon \cite{Dubovsky:2009xk}. 
Two horizontal lines show qualitatively different regimes for the effective mass: $ m_{>}$ stands for mass larger than $H_r$ and, complementarily, $m_{<}$ .  }
 \label{fig1}
 \end{center}
\end{figure}
\end{center}

A clear qualitative understanding of the dynamics of tensor modes may be arrived at by classifying them (see Fig.\ref{fig1} and \cite{Dubovsky:2009xk}) according to whether or not and to where these modes are relativistic (in the sense of satisfying $q^2/a(\tau)^2 \gg m^2_{{\rm eff}}$): 
\begin{itemize}
\item Modes relativistic at recombination belong to \textit{Region I}
\item Modes which are non-relativistic already early on, before entering the horizon, reside in \textit{Region II} 
\item Momenta entering the horizon as relativistic  but turning non-relativistic by 
 the recombination epoch populate \textit{Region III}
\end{itemize}

\noi This pictorial view refers of course to the massive modes but in the \textit{additive} regime one may reasonably expect that the exactly massless modes will generate the usual imprints to be superimposed, weighted by a relative coefficient, to the massive modes signatures.

\noi Note that the presence of the fractional relative coefficient  in front of the massive contribution to $h_{\mu\nu}$, e.g. in Eq.~(\ref{eq:mattercouplingdiag}), cannot alter the description in Fig.\ref{fig1} \footnote{What it does alter is the relative weight of the massive tensor sector in determining the gw signal.}. Indeed, the latter is obtained by judiciously comparing among each other the value over time of $q/a(t), H(t), \meff$. These quantities squared all appear in the massive tensor modes equation of motion and their relative strength, unaffected by the fractional coefficient, signals the regime one is working at (i.e. (non)relativistic, $m_{>(<)}$, inside/outside the horizon).

Let us now discuss how one may generate Fig.\ref{fig1} and then the consequences it can possibly entail for bigravity CMB signatures. Region I denotes modes relativistic at recombination. For a generic multipole moment one has, in conformal time, that $\ell \sim q(\tau_0-\tau_{{\rm r}})$. The role of the $(\tau_0-\tau_{{\rm r}})$ factor is clearly that of accounting for the time evolution to the present day.  In our setup this evolution can in principle depart  from that of $\Lambda$CDM.

Working in the LEL regime and implementing Eq.\eqref{eq:fried3}  (with $\kappa\ll1$ and $\Lambda_{{\rm eff}}\sim \Lambda = m^2 \rhomg[\xi_c]$ so that the form of the Friedmann equation reduces, formally, to the standard one) then, in multipole language the non-relativistic threshold is reached \cite{Dubovsky:2009xk}  at recombination  whenever: 
\ba
\ell \leq \meff a(\tau_{\rm r}) (\tau_0-\tau_{{\rm r}})\equiv \ell_0\sim \frac{\meff}{H_0(1+z_r)}\int^{1}_{(1+z_r)^{-1}} \frac{d a}{
{\sqrt{\Omega_{\Lambda_{{\rm eff}}} a^4 + \Omega_m a + \Omega_r }}} \sim 3.3 \frac{\meff}{H_0 (1+z_r)}\, , \nonumber \\
\label{ell0}
\ea

In the numerical calculation we have set $\Omega_{\Lambda_{{\rm eff}}}\sim  \Omega_{\Lambda} \sim 0.73$ because, as we have seen, the background evolution of the LEL regime mimics that of $\Lambda$CDM in our setup. 

\noi It is also worth pointing out that the result of a (\ref{ell0}) \textit{without} a c.c. term would amount to changing the numerical factor in front from $3.3$ to $3.7$, a mere 10\%, reflecting the fact that we owe most of the evolution since recombination to the content in $\Omega_m, \Omega_r$. 

A recombination redshift of $z_r \simeq 1088$ delivers the yellow shaded area in Fig.(\ref{fig1}). Note that for very small masses, say $m_{{\rm eff}} \lesssim 10^2 H_0$, the value of $\ell_0$ will be so small that the B-mode spectrum will be unaffected by non-relativistic modes \footnote{This is true exactly only if the contribution at reionization is neglected.  The fact that non-relativistic modes play no active role is intuitively clear upon noticing that the mass $\meff \sim 300\,H_0$ corresponds to the size of the visible universe at recombination.}. Incidentally, for $\meff=H_r$ one finds  that $\ell_0 \sim 64$, which is the only point shared by all three regions in Fig.\ref{fig1}.  

 The first, most straightforward, realization is that modes which are relativistic at recombination are entirely insensitive to the presence of the effective mass $m_{{\rm eff}}$. On the other hand, one can always probe the non-relativistic regime through modes satisfying $q/a(t\leq t_{\rm r})<m_{{\rm eff}}$. 

In the $m_{<}$ range there are only two possible configurations: modes relativistic at recombination (Region I), or non-relativistic momenta which only re-enter the horizon  after recombination (Region II). In this mass range modes are indeed forbidden from stepping in the horizon as relativistic and slowing down to non relativistic before recombination. Indeed, consider a $q/a(t<t_{{\rm r}}) > \meff$; this mode can easily transition into non-relativistic after some time. On the other hand, because $H(t)$ has a steeper time dependence than $a^{-1}(t)$ and becomes $H=H_r> \meff$ at recombination, it must be that  if the mode $q$ in question is outside the horizon it will stay out past the time it becomes non-relativistic. This dynamics then fits the description of modes populating Region II. 
The signatures of massive tensor modes in the $m_{<}$ range become distinct from those of the massless case only for long, outside the horizon, wavelengths. The source term $|\psi|^2$ which feeds the expression for the multipole coefficients depends on the time derivative of the primordial tensor perturbations $\dot{h}_{ij}$, which is in turn related to the term $q^2+\meff^2\,a(t_{{\rm r}})^2$. Clearly, at very large wavelength (from the onset of what we call the non-relativistic regime)  it is the mass contribution to dominate and provide an enhancement with respect to the massless case.  This is precisely the low-$\ell$ plateau found in \cite{Dubovsky:2009xk}. 

For completeness we report that,  as one raises the value of the effective mass, a new qualitatively different possibility emerges: relativistic modes might enter the horizon and slow down to become non-relativistic before recombination. This is the dynamics which characterizes Region III. We refer the reader to Ref.~\cite{Dubovsky:2009xk} for details on how to derive the border between Regions II and III. Our interest is focused on the modes of Region II which, as we have briefly reviewed, contribute to a plateau in the CMB tensor spectrum.

Having seen how Fig.\ref{fig1} provides an understanding on the dynamics of massive tensor modes, we pause here to note that this understanding is necessarily qualitative in nature and serves its purpose quantitatively only in the asymptotics. We have shown in detail below Eq.(\ref{ell0}) that a very small effective mass, below $10^2 H_0$, will witness most modes being relativistic and, as for the non-relativistic ones, those will not correspond to a high enough $\ell$ so as to leave any marks on the B-modes CMB.  It is safe to say that such a small effective mass would not be detected \footnote{This statement is all the more appropriate in a bigravity setup where the massive tensor modes do not have the full weight they enjoy in a purely massive gravity theory .}.

Navigating Region II for larger $\meff$ will eventually lead to the ``large wavelength outside-the-horizon" enhancement mentioned above, an effect propagating all the way to the B-modes spectrum and shaping up as a the low-$\ell$ plateau in the $C^{T}_{BB, \ell}$ multipole coefficients (for a fixed $\meff$ of this size and higher $\ell$s  one would step into Region I). 
Determining exactly the onset of this enhancement is a task best performed through the use of software such as ${\rm CAMB}$ \cite{Lewis:1999bs}, see \cite{Dubovsky:2009xk}. The result is that a plateau starts emerging at about $\meff \sim 1.2 \times 10^4 H_0$, its effect being most striking (two orders of magnitude larger than the standard massless tensor signal of GR) at $\meff \sim 1.5\times 10^5 H_0$ only to weaken and eventually become suppressed with respect to the massless signal as soon as $\meff \sim 3 \times 10^5 H_0$. 

We stress here that the plateau is a clear-cut effect that, if detected, would represent the most prominent CMBR signature of a massive theory of gravity. 
Most importantly for our analysis, the fact that this effect generically takes place in the low-$\ell$ regime will, as we shall see, guarantee that a bigravity model may also lead to such an imprint in the same effective mass range.  
Another thing to keep in mind is that in our case, as opposed to the analysis in \cite{Dubovsky:2009xk}, the unitarity bound sets a strong upper bound on $m_{\rm eff}$, of the order $H_r$. This requirement stems from the helicity-0 mode analysis and is therefore not necessarily present in e.g. Lorentz breaking theories of massive gravity that inspired the work in \cite{Dubovsky:2009xk}. 

Proceeding with the analysis at larger effective mass values one will see a suppression of the signal. The reason for the asymptotic suppression at  $\meff \gg H_r$ is that non-relativistic modes will start oscillating sooner and sooner outside the horizon with increased frequency $\meff$ leading to an averaging-out which amounts to a strong suppression of the signal. Determining exactly by what value of $\meff$ this effect will take over the enhancement is a task beyond the scope of the present work.

\para{The additive regime and the bigravity signal} The reasons Region II is of particular interest for us are manifold: besides being responsible for an intriguing low-$\ell$ plateau in massive gravity, this is an area in $(q,\meff)$-space whose contribution to multipole coefficients can be treated as ``additive" to a good approximation in the case of bigravity. 

As mentioned above, where we part ways with the work in \cite{Dubovsky:2009xk} is in considering two sets $H^{+/-}$ of tensor modes. As a consequence, our source term $|\psi|^2$ will consist also of cross terms. Crucially, in a low-$q$ and outside-the-horizon range such as Region II, the contribution of the massive tensor modes far surpasses (assuming the contributions are equally weighted) that of its massless counterpart. As a matter of principle then, not just massive gravity, but also bigravity can lead to distinct imprints in the B-modes spectrum for the appropriate $m_{\rm eff}$ range. 

In bigravity though, the nature of the signal will also depend on the value of the relative coefficient between the massive and massless tensor modes coupling to matter via e.g. Eq.(\ref{eq:diagbasis}). Schematically:
\ba
{\rm Signal} \sim \Big[(H^{+}_{\rm source})|_{m=0}+C_{rel} \,(H^{-}_{\rm source})|_{m_{\rm eff}} \Big]^2\, .
\ea

The first thing to keep in mind is that the unitarity bound immediately sets the dynamics in the $m_{\rm eff}\gtrsim H_r$ region.
A $C_{rel}\gg1$ would produce a signal profile with a very substantial overlap with the massive gravity one, with a low-$\ell$ enhancement at the specified $m_{\rm eff}$ range and a suppression in the asymptotics of a very large $m_{\rm eff}$. For $C_{rel}\sim 1$ and down to values satisfying $C_{rel}\gtrsim 1/10$ one would still be able to see the low multiple enhancement for the appropriate $m_{\rm eff}$; this is  because the maximum signal enhancement due to a mass (when $\meff \sim 3 \times 10^5 H_0$) is two orders of magnitude larger than the would-be massless signal. As for the $m_{\rm eff}>20H_r$ range in this configuration some suppression is to be expected but, again, this regime is best understood by running the software\footnote{We anticipate however that in the $m_{\rm eff}\gg H_r$ regime the source function $|\psi|^2$ might well receive an important contribution from what we call the -cross terms- of the two tensor sectors and therefore one may not rely on the massless/massive modes \textit{additivity} any longer. On the other hand, it is important to stress that a more detailed analysis of the dynamics for which cross terms play a leading role (a purely bigravity effect this one) is bound to lead to an interesting characterization of further signatures of bigravity theories. We leave this to future work.}. Finally, the $C_{rel}\ll 1/100$ configuration is bound to generate a signal with almost complete overlap with the GR profile.

In general then, the difference between a  bigravity theory and its massive gravity limit  would most clearly manifest itself as a lesser  enhancement in the low-$\ell$ plateau for the former and correspondingly a less dramatic suppression in the large $m_{{\rm eff}}$ region of the bigravity parameter space.  From the perspective of CMBR signatures one may summarize these findings as evidence that  bigravity theories generate imprints which are overall less sharp than those originating from pure massive gravity. This is justified already at an intuitive level because the bigravity theories space spans corners of pure massive gravity (e.g. the $M_f \rightarrow \infty$ limit)  but is also endowed with two additional degrees of freedom which account for the dynamics of the additional massless tensor modes. 
In principle, depending on the relative coefficient $C_{rel}$ of the $H^{+/-}$ modes contribution one might make bigravity imprints as sharp as those for massive gravity; on the other hand, those signatures can be ``watered down" towards the purely massless spectrum by a different judicious use of the same coefficient.

Zooming in the LEL regime, one can see that  $ C_{rel} \sim \mathcal{O}(1)\cdot \kappa \lesssim 10^{-5}$ thus concluding that the imprints of bigravity in the LEL are very hard to probe and the CMBR signal is expected to mimic GR.  What led to our restricted parameter space were, in addition to our working in a low-energy regime, the requirement of an almost $\Lambda$CDM background evolution combined with the constraints on the effective Planck mass. In turn, this  resulted in a very small relative weight for the massive tensor modes at recombination.  

Our setup should by no means be thought of as the only cosmologically viable option. Much more work is needed in this direction. It is indeed quite possible that another stable region may be found in the future whose domain includes a massive tensor sector generating more prominent signature in the CMB.

\section{Summary}
\label{sec:conclusion}
The observed accelerated expansion of the universe has reignited the research aimed at finding compelling theories modifying GR in the IR. 
Theories of massive gravity are a sure candidate for the part. Their phenomenology is currently under intense scrutiny. Most studies so far have focused on the late-time cosmological dynamics in these theories. In this paper we took a different approach and asked how early-time cosmology dynamics and constraints would reflect on the bigravity parameter space.

Our starting point has been a theory of bigravity, where both metrics are taken to be FLRW and matter only couples to one of them.
We studied the predictions for the tensor-to-scalar ratio, $r$, in bigravity in a specific low-energy limit. In this regime, the two copies of tensor modes, which are generally coupled, can be diagonalized in a time-independent fashion and solved for. It is then possible to write $r$ as a linear combination of contributions from the massless and massive modes. We find that in this regime, once additional constraints are imposed, the contribution from the massive tensor modes is suppressed. 

We further showed that, although the massive sector of the theory can in general leave quite a distinctive imprint on the B-modes profile, the region in the parameter space we have been probing supports a suppression of the massive tensor modes contribution to the overall signal in favour of a GR-like profile.

  We are led to conclude that a very efficient Vainshtein screening under specific\footnote{ Although, as pointed out in \cite{Lagos:2014lca}, an in depth study of the initial data problem is needed in order to decide what really is tuned and what is not.} conditions leads to a hard-to-detect, as far as CMB data is concerned, bigravity imprint. One could say that the massive tensor sector would bypass, rather than pass, the cosmic background radiation test by means of a small $\kappa$.
A posteriori this is not surprising: the assumed strong Vainshtein screens the non-tensorial d.o.f.'s and $\kappa\ll1$ effectively reduces $H^{+}_{ij}$, the massless modes, to $h_{ij}$ and suppresses $H^{-}_{ij}$. This is morally very close to GR. On the other hand, the $\kappa\ll1$ condition stems only from late time constraints on the effective Planck mass and what looks like GR at recombination may reveal itself more at solar systems scales where an analysis of the screening mechanism is more accessible via the decoupling limit.  

Further investigation is required to ascertain the degree to which an active Vainshtein can screen at the decoupling limit and, possibly, away from it. The latter task is especially interesting and important for the analysis presented here, but also especially complicated. Indeed, in our setup a small $M_f$ is expected and it corresponds to a small naive strong coupling scale $\Lambda_{3,f}\ll\Lambda_{3} $. We leave this to future work.

\para{ Relation to recent works}The phenomenology of bigravity theories has seen increased recent interest. Some works have some overlap with our own. 
Comelli et al.~\cite{Comelli:2014bqa}, Koennig et al.~\cite{Konnig:2014xva}, Lagos et al~\cite{Lagos:2014lca}, Cusin et al.~\cite{Cusin:2014psa} and Amendola et al \cite{Amendola:2015tua} employed linear perturbation theory to understand what pathologies may arise in each sector of bigravity. These works vary depending on the branch and $\alpha_n$ region they focus on but overall they span several branches of solutions including the one employed here. A common conclusion is that further investigation is needed on the initial value problem, as already stressed in \cite{Lagos:2014lca}. A step in this direction is represented by the work in \cite{Johnson:2015tfa} by Johnson et al. In particular, \cite{Johnson:2015tfa} identified a background which generates GR-like results under a large pool of initial conditions during inflation.

The low-energy regime (LEL), employed throughout  this manuscript was introduced in de Felice et al.~\cite{DeFelice:2014nja}. To the parameter space therein we
superimposed additional constraints (e.g. $\kappa \lesssim 10^{-5}$) that lead to a reduced space where the massive tensor sector is suppressed. 
In this sense then, we naturally make contact with the work by Akrami et al. in \cite{Akrami:2015qga}.

Our working in a restricted LEL regime means the scale at which one imagines initial conditions are set is far beyond the reach of our approximation and above our setup strong coupling scale so that our contact with the results in \cite{Johnson:2015tfa} is an indirect one.

While this work was nearing completion, a preprint by Sakakihara et al~\cite{Sakakihara:2015naa} appeared in which the tensor power spectrum was derived in a specific bigravity model, corresponding, in the notation of Eq.~\eqref{eq:LagMGR}, to $\alpha_{n\not= 2}=0$. The Authors considered the so-called healthy branch of solutions and, in as much as there is overlap with our work, they reached conclusions not unlike ours.
		
A number of earlier works also focused on massive gravity signatures in the CMBR. Ref.~\cite{Dubovsky:2009xk}, as we have seen, pointed out the existence of a B-modes enhancement in the low multipoles for a specific mass range. The study in Ref.~\cite{Gumrukcuoglu:2012wt} concerned a general analysis of the profile and detectability of the gravitational wave signal arising from a time-dependent mass term  in massive gravity.

	\section*{Acknowledgements}
	 It is a pleasure to thank Claudia de Rham and Andrew J.~Tolley for collaboration at early stages of this work and for illuminating conversations. We are grateful to Matthew Johnson, Adam~R.~Solomon and Alexandra Terrana for fruitful discussions. We also thank C.~de~Rham, M.~Johnson and A~R.~Solomon for very useful comments on a draft version of this paper.  The work of MF was supported in part by grants DE-SC0010600 and NSF PHY-1068380. RHR acknowledges the hospitality of  
	DAMTP at the University of Cambridge and the Perimeter Institute of Theoretical Physics.
	RHR's research was supported by a 
	Department of Energy grant de-sc0009946, 
	the Science and Technology Facilities Council grant ST/J001546/1 and in part by Perimeter Institute for Theoretical Physics. Research at Perimeter Institute is supported by the Government of Canada through Industry Canada and by the Province of Ontario through the Ministry of Economic Development \& Innovation.

\end{document}